# Proton ordering dynamics of $H_2O$ ice

Fei Yen,[1,2*] and Zhenhua Chi[1,2]

[1] Key Laboratory of Materials Physics, Institute of Solid State Physics, Hefei Institutes of Solid State Physics, Chinese Academy of Sciences, Hefei, 230031 China.
[2] High Magnetic Field Laboratory, Hefei Institutes of Physical Sciences, Chinese Academy of Sciences, Hefei, 230031 China.

**Abstract:** From high precision measurements of the complex dielectric constant of $H_2O$ ice, we identify the critical temperatures of the phase transition into and out of ice XI from ice I$h$ to occur at $T_{Ih\text{-}IX}$=58.9 K and $T_{IX\text{-}Ih}$=73.4 K. For $D_2O$, $T_{Ih\text{-}IX}$=63.7 K and $T_{IX\text{-}Ih}$=78.2 K. A triple point is identified to exist at 0.07 GPa and 73.4 K for $H_2O$ and 0.08 GPa and 78.2 K for $D_2O$ where ices I$h$, II and XI coexist. A first order phase transition with kinetic broadening associated to proton ordering dynamics is identified at 100 K.

The ice found throughout the surfaces of Earth (ice I$h$) possesses a hexagonal unit cell belonging to the space group P6$_3$/mmc.[1] Each oxygen has four hydrogen neighbours: two linked by covalent bonds which make up the $H_2O$ molecule and two weaker intermolecular hydrogen bonds from neighbouring $H_2O$ molecules. As such, each hydrogen is linked to two oxygen: one via a covalent bond and another via a hydrogen bond. These are known as the 'ice rules'.[2] From a different perspective, all the oxygen atoms can be thought to be comprised of an arrangement of tetrahedra with four hydrogen atoms residing inside. Since the position of the hydrogen proton rests closer to the covalently bonded oxygen, six different positional configurations exist inside each $H_4O_2$ tetrahedron (Fig. 1). In the ice I$h$ phase, the proton configuration inside each oxygen tetrahedron is random; while at low temperature, the proton configurations are expected to become globally ordered near 72 K to form a 'proton ordered' state. The problem however, is that localized proton hopping is limited at low temperatures which restricts proton mobility starting from about below $T_g$=136 K.[3] The freezing of such randomness of the covalent and hydrogen bonds is the reason why a residual entropy exists at absolute zero.[4,5] The ground state of one of the most abundant solids in the universe remains inaccessible in laboratories because its full natural formation can take up to 100,000 years.[6]

To speed up the formation process of proton ordered ice, doping $H_2O$ with KOH (which helps the mobility of the $H_2O$ molecules) and maintaining it below 72 K for days has allowed for the synthesis of domains of ice XI,[7] the proton-ordered form of ice I$h$. Ice XI was subsequently found to be orthorhombic (space group $Cmc2_1$).[8] The ratio of the ice XI to ice I$h$ mixture however, depends on the annealing time and dopant ratio[9] which leads us to believe that an infinitesimal amount of ice XI must nucleate in pure $H_2O$ at the start of each process. However, the amount of ice XI initially formed in pure $H_2O$ is so small that it is not readily detectable by spectroscopy and calorimetric measurements because it is superimposed by a large ice I$h$ background. On the other hand, a continuous measurement of the complex dielectric constant while varying the temperature can detect *changes* that directly stem from the protons (since they possess a dipole) as well as any nucleation of ice XI domains within a static ice I*h* matrix background. Hence, the ice XI phase transition in pure $H_2O$ as well as any associated proton ordering dynamics should be reflected in the dielectric constant.

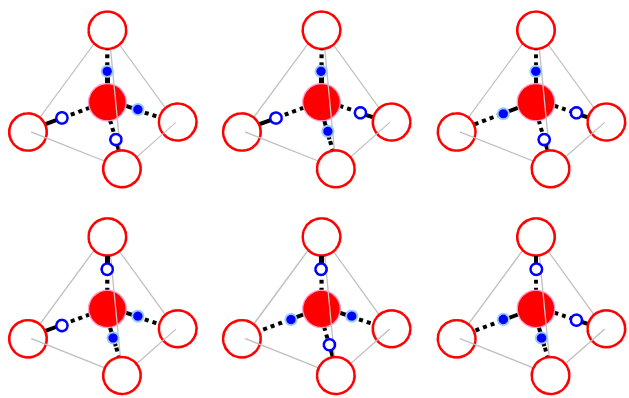

**Fig. 1** The six possible configurations an ice structure can take within an $H_4O_2$ tetrahedron. Red and blue spheres are, respectively, oxygen and hydrogen atoms. Covalent bonds are drawn as lines (—) and hydrogen bonds as dotted lines (•••). The $H_2O$ molecule in each panel is represented by filled spheres. The proton ordered state occurs when all tetrahedra possess the same proton configuration globally.

In this paper, we present a study on isobaric scans of the complex dielectric constant of specially prepared samples of solvent-free $H_2O$ and $D_2O$. We detect for the first time, the presence of ice XI domains in pure $H_2O$ and $D_2O$. From such, we identify the phase boundaries between ices XI and I*h* during cooling and warming; identify the stability range of the ice XI phase in the pressure domain; propose a probable mechanism underlying the continuous proton ordering and disordering processes; and prove the existence of a triple point where ices I*h*, II and XI coexist.

Ice XI domains were obtained by compressing deionized and degassed liquid $H_2O$ up to 0.03, 0.05, 0.07 and 0.08 GPa at room temperature and cooled to ~20 K. The cooling rates were 3-5 K/min.

During cooling, ice I$h$ formed first, typically around 256 K from supercooled water, before ice XI domains began to nucleate below 72 K. Pressure was applied by a beryllium copper double piston clamp cell. Variation of temperature was enabled by a customized gas exchange cryostat. The real ε'($T$) and imaginary parts ε"($T$) of the dielectric constant were obtained by measuring the capacitance and loss tangent, respectively, of a pair of platinum parallel plate electrodes inside the sample space at 1 kHz with an Andeen Hagerleen (AH2500A) ultra-precision capacitance bridge with 1 attoFarad resolution. The area of the electrodes was 1.5 x 4.0 mm and 0.5 mm apart from each other. This allows us to detect changes in the dielectric constant that are in the order of ~$10^{-15}$. The pressure medium was also $H_2O$ so the electrodes were surrounded by $H_2O$ all around with the measured ε'($T$) and ε"($T$) coming from the active area in between the plate electrodes. Pressure was also only applied whenever the sample was in liquid form to make sure the sample was under hydrostatic pressure. These steps allowed us to probe the dielectric properties of large crystals minimal of shear stresses and surface effects which are of vital importance since strain energy inhibits transformation into the ice XI phase according to calculations of the Gibbs energy.[10] The pressure inside the sample space was determined from the temperature of the ice I$h$ liquidus line indicated by the sudden increase of ε'($T$) and ε"($T$) by a couple orders of magnitude during heating.

For the formation of D$_2$O ice XI domains, the preparation and experimental processes were the same as H$_2$O, except that 99.9% isotopic purity D$_2$O acquired from Sigma-Aldrich was used.

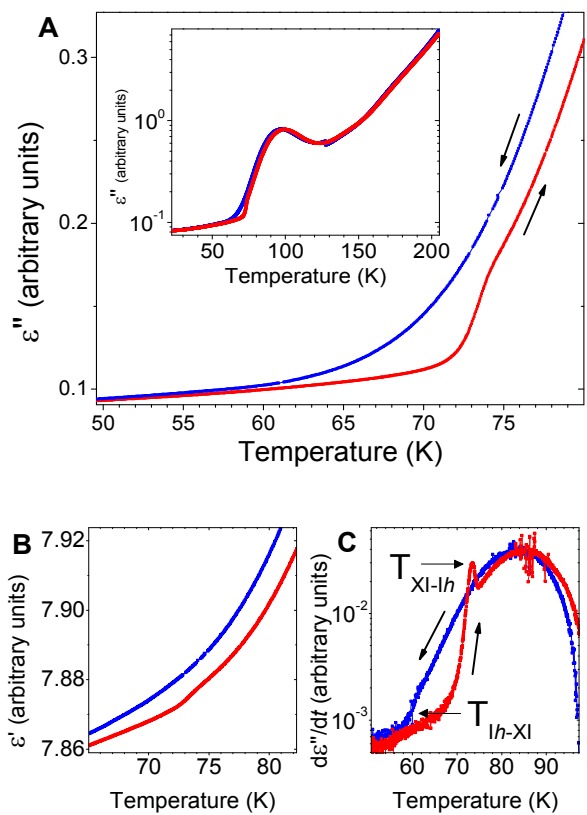

**Fig. 2** Complex dielectric constant of H$_2$O at 0.05 GPa. (A) Imaginary part of the dielectric constant as a function of temperature ε"(T) where a clear anomaly can be observed near 74.1 K. Inset: ε"(T) in a larger scale showing continuous hydrogen bond (dis)ordering occurring in the 59 < T < 124 K range. (B) Real part of the dielectric constant showing a less pronounced anomaly at 72.7 K. (C) First order derivative of ε"(T). Peak at $T_{XI-Ih}$=73.4 K during warming represents the phase transition of ice XI to ice Ih. Change in slope at $T_{XI-Ih}$=58.9 K during cooling is the temperature where ice XI domains begin to nucleate from ice Ih.

Figure 2(A) shows the cooling and warming curves of the imaginary part of the dielectric constant with respect to time ε"(T) of H$_2$O at 0.05 GPa. During warming, a continuous uptick occurred at 72.4 K followed by a change in slope at 74.1 K. The real part of the dielectric constant ε'(T) also showed anomalies at the same temperatures, though less pronounced

(Fig. 2(B)). Fig. 2(C) shows the first order derivative of $\varepsilon''(T)$ showing a clear peak anomaly at $T_{XI-Ih}$ = 73.4 K representing the phase transition from ice XI into ice I$h$. This result is in good agreement with Kawada's dielectric constant measurements taken at lower frequency for KOH-doped ice,[11] although they reported observing a phase transition only upon warming. In contrast to our case, from Fig. 2(C), a clear discontinuity at $T_{Ih-XI}$ = 58.9 K was also observed during cooling indicating the phase transitioning into ice XI from ice I$h$. Unfortunately, Kawada only presented data down to 67 K and 71 K for KOH-doped H$_2$O (Ref. 11) and KOD-doped D$_2$O (Ref. 12) ices, respectively, so $T_{Ih-XI}$ was not reported in their studies. The discontinuities found in the first order derivatives of $\varepsilon''(T)$ and hysteresis of 6 K confirms the first order type phase transition in and out of ice XI. Neutron powder diffraction[13] and calorimetric studies[14] on doped samples have also claimed the first order nature of $T_{XI-Ih}$.

Figure 3 shows warming curves of $\varepsilon''(T)$ at 0.03, 0.05, 0.07 and 0.08 GPa. Two runs were performed at 0.03 GPa. The magnitude of $\varepsilon''(T)$ for the second run between 76<$T$<132 K was higher than the first run. This can be explained by results obtained from a neutron powder diffraction study[13] claiming that a small portion of the proton ordered domains is retained in the ice I$h$ phase when warming from ice XI. Above 0.07 GPa, both $T_{XI-Ih}$ and $T_{Ih-XI}$ as well as the maximum near 100 K and minimum at

124 K were no longer present indicating the collapse of the ice XI phase and its associated proton dynamics beyond this pressure.

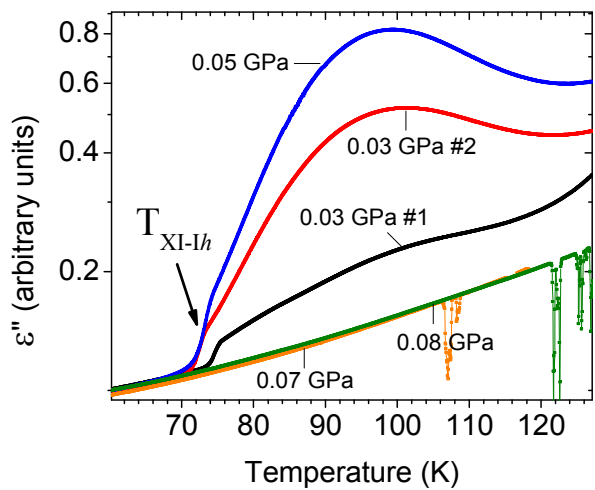

**Fig. 3** $\varepsilon''(T)$ of $H_2O$ during warming at different pressure. Above 0.07 GPa, $T_{XI-Ih}$ was no longer observed. Second run at 0.03 GPa exhibited a higher amplitude compared to the first run since some ordered hydrogen bond domains were retained from the first run.

In quenched disordered systems which usually involve the cooling of a system that possesses defects or impurities, large regions of hysteresis occur at low temperature.[14] Kinetic broadening of a peak anomaly also occurs if a first order phase transition exists. In the current case, the protons are frozen-in in a disordered fashion so ice can also be considered to be a quenched disordered system. From such, we conjecture that the maximum in $\varepsilon''(T)$ near 100 K is a broadened first order phase transition (Figs. 2(A) and 3) strictly associated to proton ordering. The observed continuous proton ordering and disordering near 100 K is a manifestation of the protons (not the oxygen atoms) being frozen into a glassy state during cooling. This also explains why a large hysteretic region exists between $T_{XI-Ih}$ and $T_{Ih-XI}$.

Figure 4 shows the warming curves of $\varepsilon''(T)$ at different pressure for pure $D_2O$. The respective critical temperatures were on average 4.5 K higher than $H_2O$, but expected since the transition lines in $D_2O$ are typically 4 degrees higher than $H_2O$ (Ref. 15) due to a higher zero point energy. $T_{XI-Ih}$ was observed at 0.07 GPa but no longer at 0.08 GPa placing the stability region of $D_2O$ ice XI less than 0.01 GPa higher than $H_2O$ ice XI in the pressure domain.

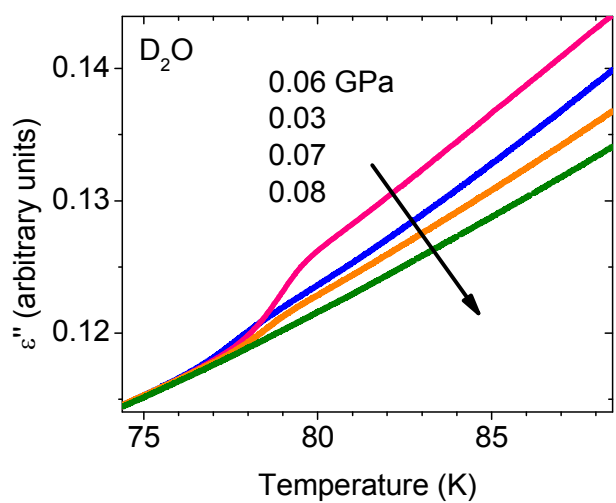

**Fig. 4** $\varepsilon''(T)$ of $D_2O$ at different pressure. $T_{XI-Ih}$ was on average 78.2 K up to 0.70 GPa. Above 0.80 GPa, $T_{XI-Ih}$ was no longer detected.

Figure 5 shows the stability regions of ice XI with respect to ice I$h$ for $H_2O$ and $D_2O$. This marks the first time where the $T_{XI-Ih}$ and $T_{Ih-XI}$ phase transition lines have been determined experimentally in pure $H_2O$ and $D_2O$. In all other cases, some type of catalyst dopant has been employed. Amazingly, the extrapolated ice I$h$/II phase line (dashed line in Fig. 5)[17] to low temperature passes exactly through 0.07 GPa and 73 K suggesting that ices I$h$, II and XI coexist at this point, the existence of a triple point. According to the 180° rule,[16] a metastable extension of the $T_{XI-Ih}$ phase line can exist beyond the triple point (dotted line in Fig. 5). This means that beyond 0.07 GPa, only ice XI can phase transition to ice I$h$ but not the other way around. Metastable ice XI can still exist beyond 0.07 GPa as long as pressure is applied at $T<T_{Ih\text{-}XI}$ avoiding the metastable phase line that forbids the formation of ice XI. Under such conditions, $T_{XI\text{-}Ih}$ becomes observable upon heating when $P>0.07$ GPa. Indeed, calorimetric[18] and thermal conductivity[19] studies on KOH-doped $H_2O$, where pressure was applied at $T<T_{XI\text{-}Ih}$, have reported ice XI phase transitioning to ice I$h$ up to 0.16 GPa (diamond in Fig. 5).

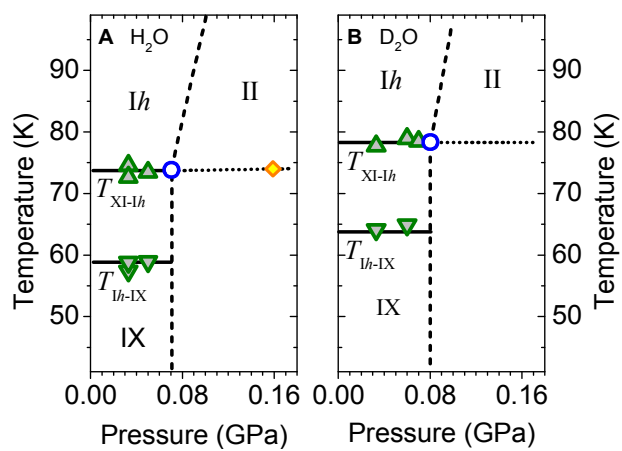

**Fig. 5** Low pressure low temperature region of the phase diagram for (A) $H_2O$ and (B) $D_2O$. Apart from $T_{XI\text{-}Ih}$ (triangles pointing up), the transition into the ice XI phase at $T_{Ih\text{-}XI}$= 58.9 K for $H_2O$ and 63.7 K for $D_2O$ is also displayed as triangles pointing down. The ice I$h$/II phase line was linearly extrapolated from high temperature (dashed line) and found to intersect with $T_{XI\text{-}Ih}$ exactly at 0.07 GPa and 73.4 K for $H_2O$ forming a triple point (blue circle) where ices I$h$, II and XI coexist. For $D_2O$, the ice I$h$/II/XI triple point occurs at 0.08 GPa and 78.2 K. Vertical dashed line represents the pressure beyond which metastable ice XI may form. Dotted line is a metastable extension of the $T_{XI\text{-}Ih}$ phase line into the phase space of ice II. The data point represented by a diamond is after Ref. [18].

**Conclusions**

We synthesized large $H_2O$ and $D_2O$ crystals under hydrostatic pressure with embedded platinum electrodes which allowed us to perform highly sensitive measurements of the dielectric constant. From such, we detected the critical temperatures of the phase transition into and out of ice XI from ice I$h$. The stable phase of ice XI resides below 73 K and 0.07 GPa for $H_2O$ and below 78 K and 0.08 GPa for $D_2O$. However, metastable ice XI may also exist at slightly higher pressures if the sample is pressurized at low temperature. Based on the 180º rule, a triple point is proven to exist where ices I$h$, II and XI coexist. The continuous hydrogen bond ordering and disordering dynamics reflected in $\varepsilon''(T)$ near 100 K is attributed to the protons being quenched in a disordered fashion at a higher temperature. Our experimental method can be applied to study the phase diagrams of other molecular solids in more detail such as CO, $NH_3$, or $CH_3$, which are also abundant in our universe.

Our work was made possible in part via the support of the National Natural Science Foundation of China grants No. 11374307, No. 51372249, and the Director Grants of the Hefei Institutes of Physical Science, Chinese Academy of Sciences.